\documentclass[secnumarabic,
               amssymb, 
               nobibnotes, 
               reprint,
               pra]{revtex4-2} 
               
\usepackage{soul} 

\usepackage{hyperref}  
\usepackage{multirow}  
\usepackage{subfigure} 
\usepackage{array,booktabs} 
\usepackage{siunitx}

\usepackage{bbm}       
\usepackage{bm}        
\usepackage{mathtools} 
\usepackage{braket}    

\usepackage[T1]{fontenc}
\usepackage{xcolor} 
 
\definecolor{themeBlue}{RGB}{38,70,83}
\definecolor{themeGreen}{RGB}{42,157,143}
\definecolor{themeYellow}{RGB}{233,196,106}
\definecolor{themeOrange}{RGB}{244,162,97}
\definecolor{themeRed}{RGB}{221,101,81}

\hypersetup{
      breaklinks=true,  
      colorlinks=true,
      urlcolor=themeBlue,
      linkcolor=themeRed,
      citecolor=themeGreen,
} 

\setstcolor{themeRed}

\begin{document}

\title{Quench dynamics in the Jaynes-Cummings-Hubbard and Dicke models}%

\author{Andrew R. Hogan}%
\email{arhogan@student.unimelb.edu.au}
\affiliation{School of Physics, University of Melbourne, Parkville, 3010, Australia}
\author{Andy M. Martin}
\affiliation{School of Physics, University of Melbourne, Parkville, 3010, Australia}
\date{Feb 2022}%
\begin{abstract}
Both the Jaynes-Cummings-Hubbard (JCH) and Dicke models can be thought of as idealised models of a quantum battery. In this paper we numerically investigate the charging properties of both of these models. The two models differ in how the two-level systems are contained in cavities. In the Dicke model, the $N$ two-level systems are contained in a single cavity, while in the JCH model the two-level systems each have their own cavity and are able to pass photons between them. In each of these models we consider a scenario where the two-level systems start in the ground state and the coupling parameter between the photon and the two-level systems is quenched. Each of these models display a maximum charging power that scales with the size of the battery $N$ and no super charging was found. Charging power also scales with the square root of the average number of photons per two-level system $m$ for both models. Finally, in the JCH model, the power was found to charge inversely with the photon-cavity coupling $\kappa$.
\end{abstract}
\maketitle

\section{Introduction}
Energy storage capabilities and efficiency by electrochemical batteries have rapidly improved in recent times, pushed by the need to robustly deal with the ever increasing energy demands of daily life. As we advance technologically in the search for faster charging batteries, recently the idea of a quantum battery has become a more heavily researched topic \cite{Quach_2022_SA, Gem_2022_IBM, Jar_2016_NJP, Ali_2013_PRE, Bin_2015_NJP, Hov_2013_APS, Kon_2022_APS, Gem_2022_BAT, San_2021_APS, Lev_2018_APS, Dou_2022_APS, Crescente_2020, e23050612, doi:10.1021/acs.jpcc.9b06373, PhysRevResearch.2.023113, PhysRevB.99.205437, PhysRevB.100.115142, Cruz_2022, PhysRevLett.111.240401, PhysRevLett.122.047702, PhysRevResearch.2.023095, PhysRevA.103.033715, PhysRevA.104.043706, PhysRevA.104.L030402}. The goal underpinning the exploration of a battery made of single quantum bits each with a single excited state is to use quantum phenomena to engineer a greatly improved energy storage device. Some limiting factors for classical electrochemical batteries are their thermodynamic energy loss due to heat and their increasing charging times for scaled up batteries \cite{Usdin_2015_PRX, Bha_2021_PJB, Skrzypczyk2014, https://doi.org/10.48550/arxiv.1812.10139}. Investigating ways that a quantum battery can deal with these issues has lead to the desire to understand how quantum states might be utilised to produce a battery with minimal energy loss and how the system can be built to minimise its charging time \cite{Ferraro_2018_PRL, Hu_2021_arx, Hu_2022_IOP, Cam_2017_APS, PhysRevB.98.205423, Friis2018precisionwork, PhysRevLett.125.040601, PhysRevLett.125.236402, PhysRevResearch.4.013172, PhysRevE.99.052106, Chang_2021, PhysRevE.103.042118}. 

Previous theoretical work \cite{Ferraro_2018_PRL} found that quantum batteries can display a super-charging characteristic. They found that as the number of two-level systems ($N$) in the battery increased, the speed with which the battery charged increased at a rate of $N\sqrt{N}$. This result has ignited significant interest in quantum batteries and inspired us to explore quantum batteries in the context of the Dicke model \cite{Dicke_1954_APS} and the Jaynes-Cummings-Hubbard (JCH) model \cite{Jaynes_1963}. 

Functionally, a quantum battery can be thought of as idealised two-level system inside a cavity whose mode is able to excite the two-level system. For such a system the battery can be thought of as being charged (uncharged) when the two-level system is in the excited (ground) state. Figure 1 schematically describes the two systems we will consider in this work. Specifically the JCH model, Fig. 1(a) and the Dicke model, Fig. 1(b), under the charging protocol shown in Fig 1(c).
In each case we sonsider a scenario where we have $N$ elements in the quantum battery. The system is initialised such that the two-level systems are in the ground state.  At $t=0$ the coupling between the two-level systems and the photons is quenched from $0$ to $\beta$. We will first consider the charging in the JCH model in sections \ref{sec:jch_theory} \& \ref{sec:JCHresults}. For the JCH system we find that the maximum charging power, $P_{max}$, is proportional to the number of the cavities in the JCH system. Additionally, we find that the maximum charging power is (inversely) proportional to square root of the number photons initially in each cavity (the photon coupling between individual cavities). The result that the maximum charging power is proportional to the number of two-level systems in the JCH model then prompts us to revisit, in sections \ref{sec:Dicke_theory} \& \ref{sec:dicke_results}, results for the Dicke model, where we construct the Dicke Hamiltonian to ensure that the thermodynamic limit is bounded. For such a regime we regain a scaling for $P_{max}$ proportional to the number of two-level systems in the Dicke cavity. 
\begin{figure}[h]
	\centering
	\includegraphics[width=1.05\linewidth]{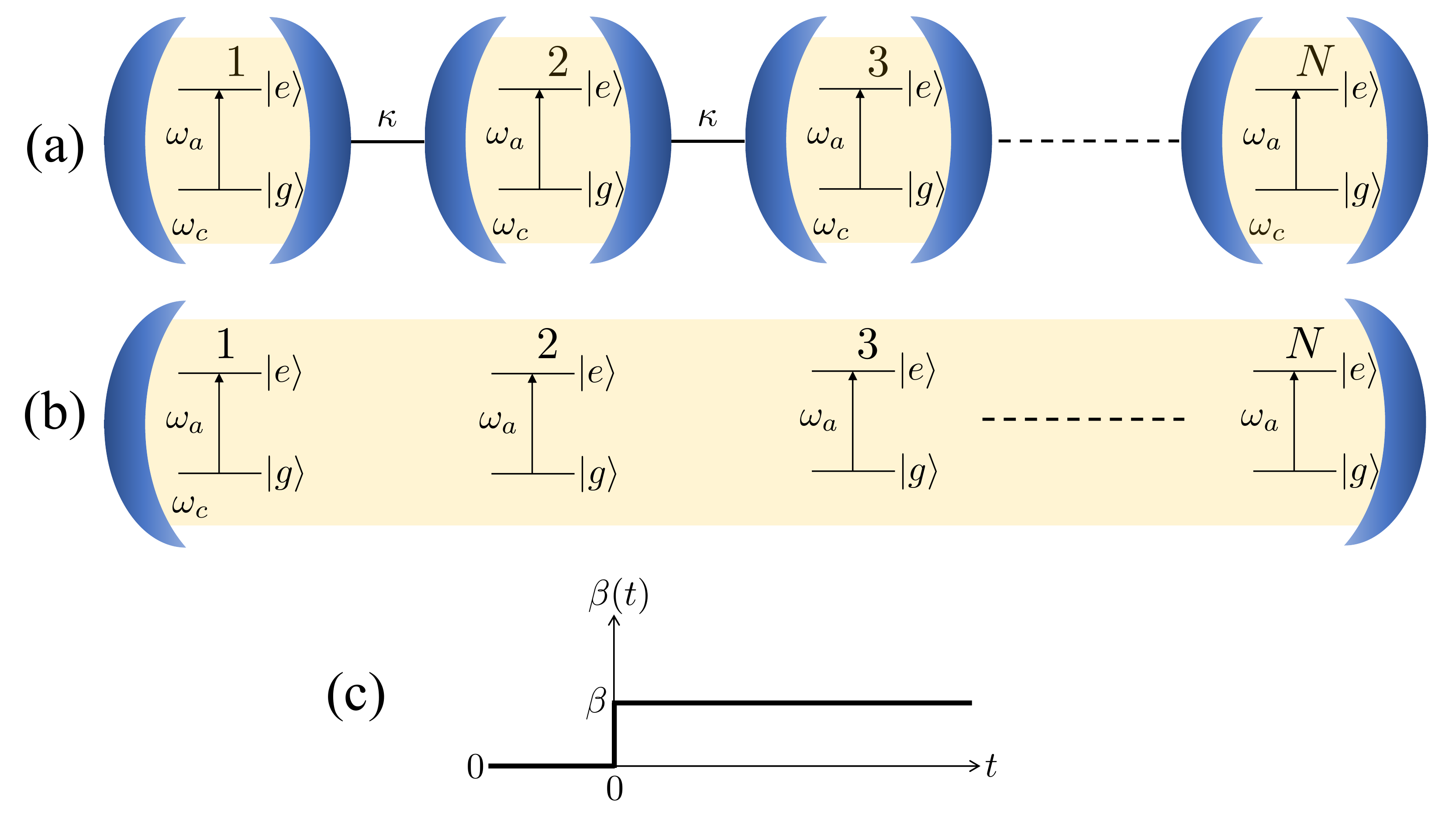}
	\caption{(a) Schematic for the JCH model. $N$ identical two-level systems each occupying their own cavity, with photons coupling between cavities with strength $\kappa$. (b) Schematic for the Dicke model. Two-level systems the same as above except that they are all in the one cavity. (c) Representation of the charging sequence of the quantum battery. Initially the photon coupling to the two-level system $\beta$ is zero, then it is quenched to a value $\beta>0$, where charging begins.} 
	\label{fig:jch_1}
\end{figure}

\section{JCH quantum batteries}\label{sec:jch_theory}

The JCH model can be thought of as representing an atom with a single excited state in the presence of $n$ photons inside a cavity. The two-level atomic system is coupled to the photons in the cavity via $\beta$, and the photons with frequency $\omega_c$ are coupled between the $N$ identical cavities via $\kappa$. Specifically the JCH Hamiltonian \cite{Kirton_2019_AQT} is ($\hbar =1$)

\begin{align}
H_{JCH}=&\sum_{n=1}^N \omega_ca^\dagger_na_n + \sum_{n=1}^N\omega_a\sigma^+_n\sigma^-_n
\notag+\beta \sum_{n=1}^N(a_n\sigma^\dagger_n+a^\dagger_n\sigma^-_n)\\
&-\kappa  \sum_{n=1}^N(a^\dagger_{n+1}a_n+a^\dagger_na_{n+1}) \label{eq:JCH_ham}
\end{align}
where $\omega_a$ is the energy of separation between the energy levels of the TLS, $a^\dagger$ and $a$ are the photonic raising and lowering operators, and $\sigma^+$ and $\sigma$ are the spin raising and lowering operators. 

Diagonalising the JCH Hamiltonian allows the Time Dependent Schrodinger Equation (TISE) to be solved and the dynamics analysed. Starting with the system in the lowest energy eigenstate, the atom-photon coupling is quenched from $\beta=0$ to $\beta>0$ at time $t=0$. In doing so, the two-level systems are taken from a parameter space where they cannot charge, and instantaneously quenched to one where they are able to begin charging. In order to quantify the charging rate we define that the energy of the system is the difference between the energy of the time varying energy and that of the initial state,

\begin{equation}
    E_\beta(t)=\omega_c\{ \braket{\psi^N_\beta(t)|\hat{J}_z|\psi^N_\beta(t)}-\braket{\psi^N(0)|\hat{J}_z|\psi^N(0)} \} , \label{eq:JCH_energy}
\end{equation}
where the energy operator for the atomic spin is
\begin{equation}
    \hat{J}_z = \omega_a \sum_{n=1}^N \sigma^+_n\sigma^-_n .
\end{equation}
With the time varying energy we find the maximum charging power of the battery by taking the maximum rate of change of the energy with respect to time,

\begin{equation} P_{max} = {\rm max}\bigg[\frac{E_{\beta}(t) }{t}\bigg], \label{eq:pmax}\end{equation}
which has a charging time to reach $P_{max}$ of $\tau$. This definition of power has been used to make a direct comparison with with existing literature \cite{Ferraro_2018_PRL}. Alternatively, the time to charge the battery to its maximum energy was explored, with both methods returning results with the same scaling factors. The two ways to analyse the power of the quantum battery are to consider how long it takes to fully charge the battery, which has a strong analogous relationship between classical batteries, or to consider the best possible charging power and consider how that scales. In the rest of this paper we will use the later definition, as in equation \eqref{eq:pmax}.  

The limit $\kappa=0$ represents the case where individual elements of the cavities are not coupled to each other, and there is no photon transfer between them. We will use this as the baseline by which we analyse how different parameters may change the charging rate of the battery, with express interest in whether increasing the size of the battery improves the charging power. With a system of isolated ($\kappa=0$) JCH two-level systems, the behaviour reduces to that of individual Rabi two-level systems with Hamiltonian,

\begin{align}
H^{JCH}_m =  
\begin{pmatrix}
m+\Delta & \sqrt{m}\beta \\
\sqrt{m}\beta & m\omega \label{eq:JCH_singleTLS}
\end{pmatrix}
\end{align}
where $\Delta=\omega_a-\omega_c$ and the average number of photons per two-level system is $m$. In this regime the JCH model can be solved analytically and has its first maximum energy at time 
\begin{equation}
\tau = \frac{\pi}{2\Omega},\label{eq:Rabi_TLStau}
\end{equation}
where the Rabi frequency is 
\begin{equation}
\Omega = \frac{\sqrt{\Delta^2+4m\beta^2}}{2}.
\end{equation}
It can be seen from equation \eqref{eq:Rabi_TLStau} that when the energy separation between the two energy levels and the photon mode energy is zero ($\Delta = 0$), the charging time will scale with the number of photons according to $\tau \propto 1/\sqrt{m}$, and $E$ scales proportional to $N$. It follows that $P_{max} \propto N$ and $P_{max}\propto\sqrt{m}$. It is therefore of interest to explore how this relationship changes when the two-level systems are able to interact. Allowing the cavities in the quantum battery to interact via photon coupling ($\kappa>0$) makes it possible to analyse how $\kappa$, $N$ and $m$ effect it's charging power.


\section{JCH Results} \label{sec:JCHresults}

\begin{figure}[h]
	\centering
	\includegraphics[width=0.85\linewidth]{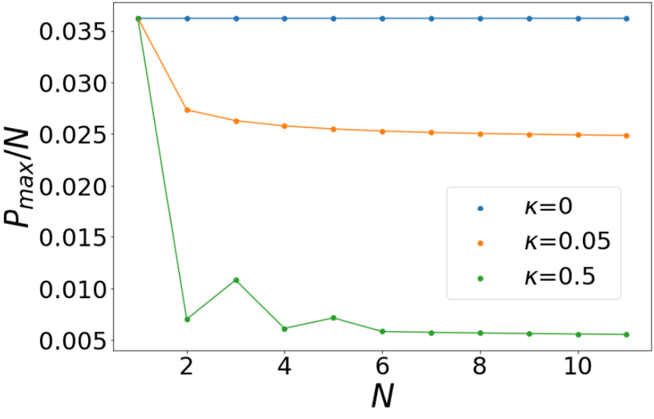}
	\caption{Charging power of the JCH quantum battery. Power is scaled with a factor of $1/N$ and plotted as a function $N$. Each line corresponds to a decreasing value of $\kappa$, with the uncoupled cavities having the largest values for the scaled power. Here $\beta=0.05$ so that $\kappa$ varies from below its energy scale to larger by an order of magnitude. The average number of photons per cavity is set to $m=1$.  The initial state is set as the state in which each cavity has $ m$ photons and is in the ground state. }
	\label{fig:jch_N_1}
\end{figure}

In this paper we present results in natural units where $\hbar=1$, and for a resonant regime where the dimensionless photon mode energy and the dimensionless atomic energy separation are both 1, and hence $\Delta=0$. In Fig.\ref{fig:jch_N_1}, the effect of increasing battery size is shown for different values of the photon mode coupling parameter $\kappa$. When $\kappa=0$, the JCH model has an analytical solution. The present simulation results overlap exactly with the analytical results obtained from the Rabi matrix of equation \eqref{eq:Rabi_TLStau}. This serves as the starting point for the comparison of the power for larger JCH systems. It can be seen in this figure, that the charging power of the quantum battery for any value of $\kappa$ never exceeds that of the completely uncoupled $\kappa=0$ case. The quantum battery has the largest maximal charging power when it acts as if it was $N$ independent single atom batteries. With the initial state taken as the lowest energy eigenstate, increasing $\kappa$ moves the state to higher energy eigenstates faster. This appears to decrease the charging power of the quantum battery. 
The maximum charging power of the JCH quantum battery was scaled by a factor of $1/N$ and for each value of $\kappa$, the data tends towards a constant. This strongly implies that $P_{max} \propto N$, the result obtained for uncoupled ($\kappa=0$) JCH two-level systems. There is a notable difference between the $P_{max}$ of odd and even numbers of cavities, as $\kappa$ increases to values larger than $\beta$. The data points show this alternating behaviour for $\kappa=0.5$, but it can be seen that this has no effect on the large $N$ behaviour of the quantum battery. 


\begin{figure}[h]
	\centering
	\includegraphics[width=0.85\linewidth]{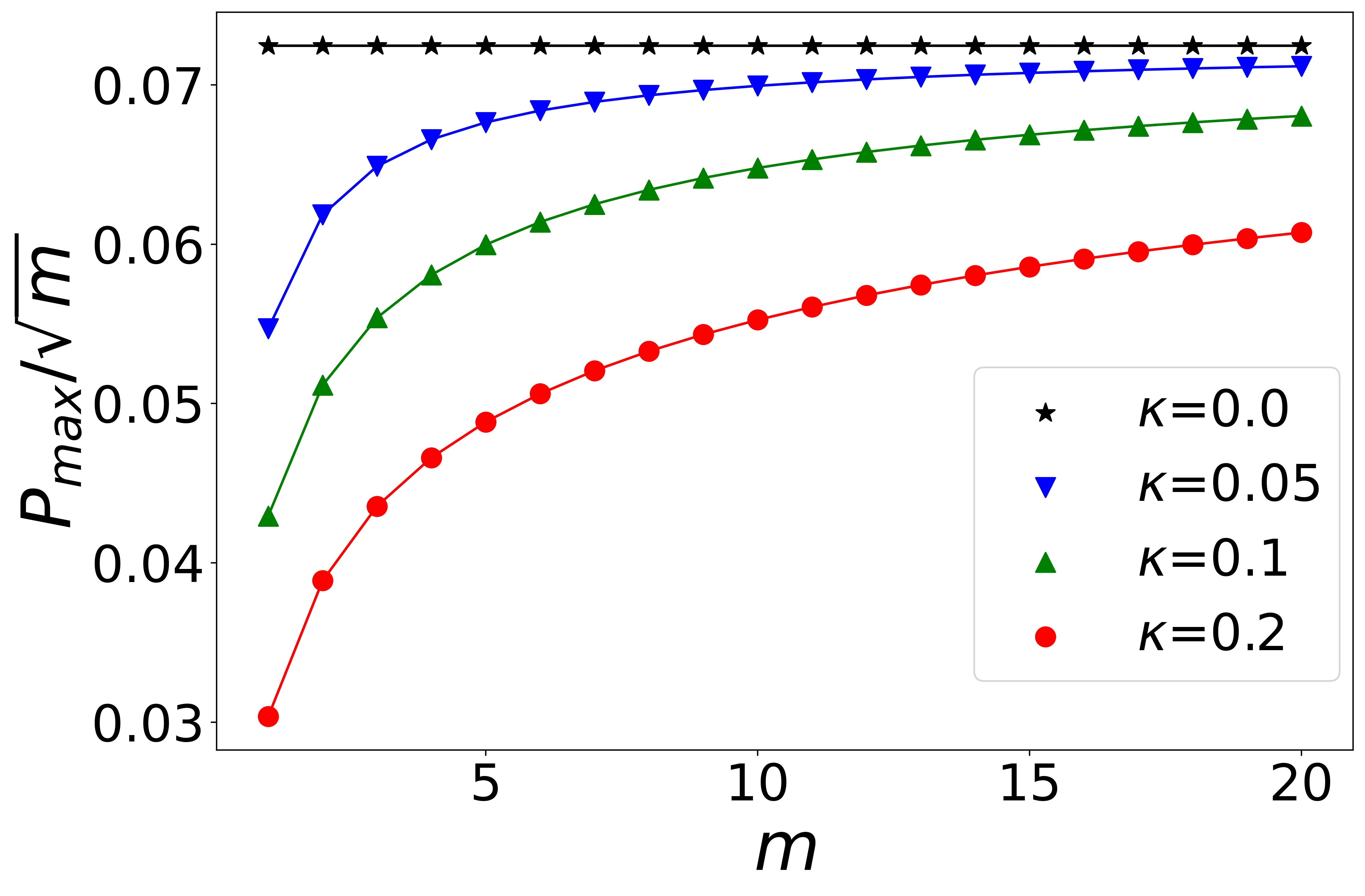}
	\includegraphics[width=0.85\linewidth]{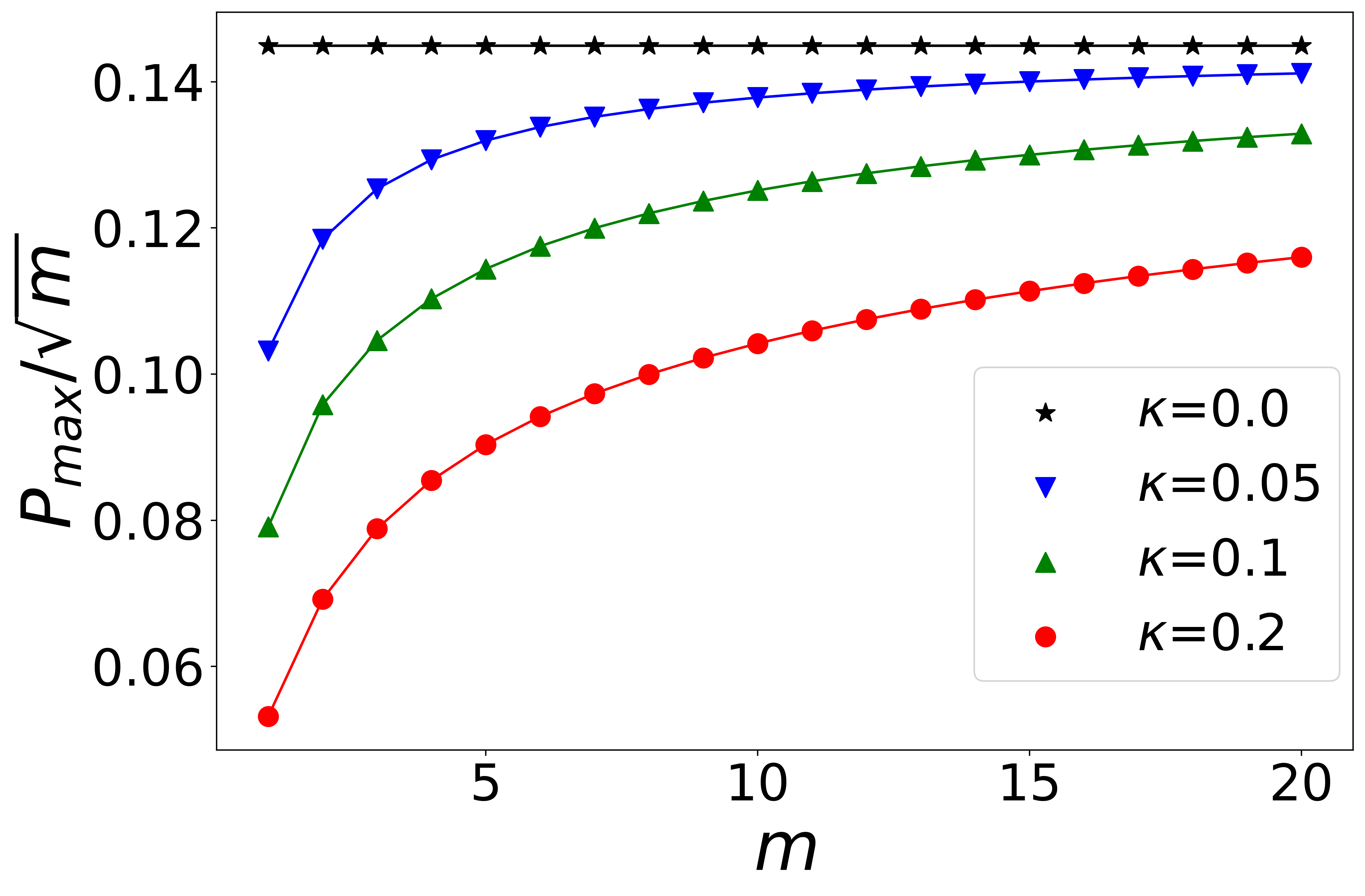}
	\caption{Number of photons per two-level system at $t=0$ vs the scaled power of the quantum battery for 2 (a) and 4 (b) two-level systems. The maximum power scaled with a factor of $1/\sqrt{m}$ to highlight that it tends towards a constant value as $m$ increases, implying that the power scales with the $\sqrt{m}$. Larger values of $\kappa$ show decreases power but appear to require larger values of $m$ from them to approach a constant value.}
	\label{fig:jch_m}
\end{figure}

Delving into the JCH quantum battery further, looking for other methods of improving their power scaling, Fig.\ref{fig:jch_m} highlights the effect that increasing the average number of photons per cavity has on $P_{max}$, scaled by $\frac{1}{\sqrt{m}}$. In Fig.\ref{fig:jch_m} (upper), there are 2 cavities with a varying number of photons $m$, at $t=0$, and it can be seen that the scaled power tends towards a constant as $m$ increases, strongly implying that the power scales as
\begin{equation}
    P_{max}\propto\sqrt{m}.
\end{equation}
This relationship can also be seen in with 4 cavities (Fig.\ref{fig:jch_m} (lower)). The same relationship is observed for up to 6 cavities, with the limiting factor being that as the number of cavities increases the size of the Hilbert space quickly makes the diagonalisation of the Hamiltonian computationally intensive. As a result, it can be seen for $N=2$ there are solutions for relatively large $m$, while $m$ has to be limited for $N>2$.   \\
\begin{figure}[h]
	\centering
	\includegraphics[width=0.85\linewidth]{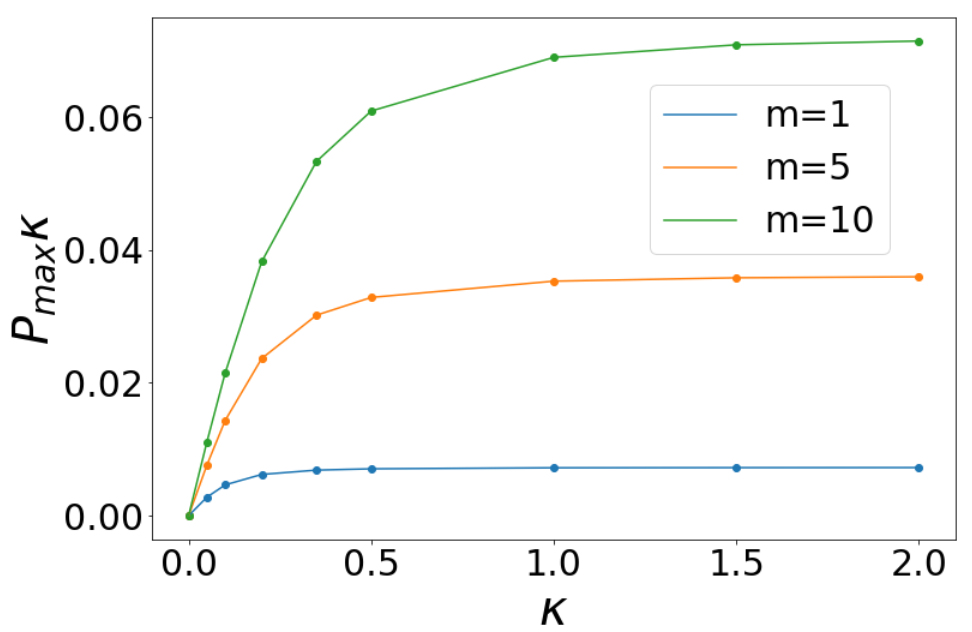}
	\includegraphics[width=0.85\linewidth]{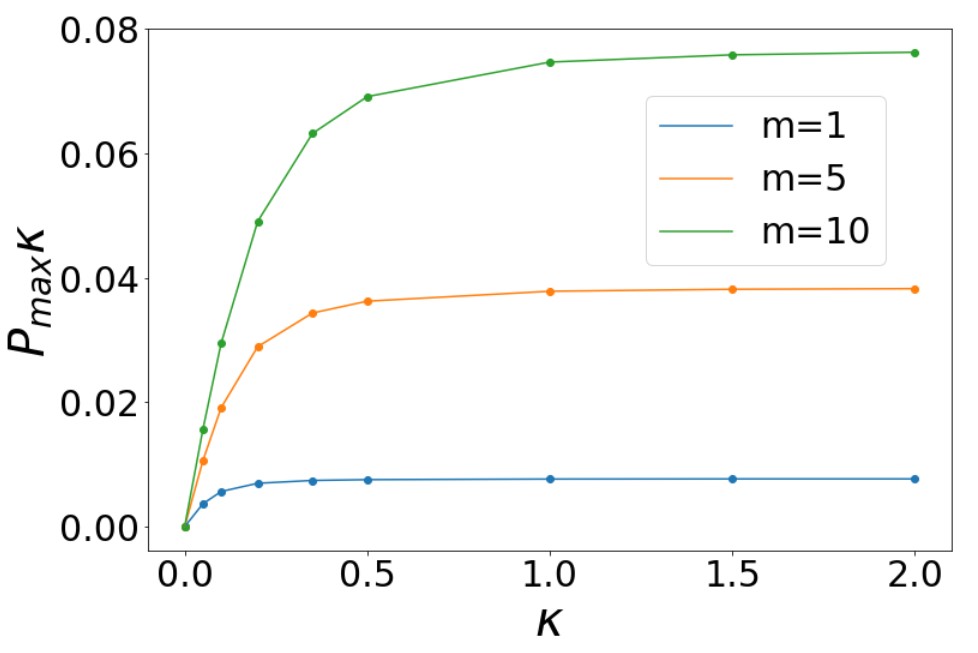}
	\caption{The maximum power of a JCH quantum battery scaled with a factor of the photon coupling $\kappa$ as a function $\kappa$. (a) The upper plot show the behaviour of $N=2$ two-level systems, while (b) the lower shows $N=3$.The tendency of this data towards a straight line for increasing values of $\kappa$ for both $N=2$ and $N=3$ cavities helps illuminate that the power scales inversely with $\kappa$. $\kappa$ covers the regime where the two-level systems are completely uncoupled ($\kappa=0$) to when it is the equal in strength to the photon/two-level system coupling strength ($\beta=\kappa=0.05$). Finally $\kappa$ ids increased to where it is the dominant energy scale of the system ($\kappa>>\beta$). } 
	\label{fig:jch_k}
\end{figure}

Another factor in the power scaling of the JCH quantum battery is the strength of the photon coupling between adjacent cavities, $\kappa$. Figure \ref{fig:jch_k} demonstrates that $\kappa$ has an inverse scaling relationship with the maximum power, by showing that the power, scaled by $\kappa$ becomes a constant for as $\kappa$ increases, highlighting that, 
\begin{equation}
    P_{max}\propto\frac{1}{\kappa}.
\end{equation}
Most interesting about this finding, is that the closest relationship to be drawn between the JCH and Dicke models are for high values of $\kappa$. This is due to the Dicke model considering each two-level system as indistinguishable from one another and are all sharing the same photons inside the one cavity. One would expect that for values of $\kappa$ where it is the dominant effect in the system, where $\beta=0.05$ and $\kappa>0.5$ the type of super scaling that was seen in \cite{Ferraro_2018_PRL} would start to show an effect.

In this paper we consider photon coupling between nearest neighbours in a line configuration. As an aside, the hopping of photons between any other cavity (hyper hopping) in the system is also explored for mathematical interest as well as closer comparison to the Dicke model, where each of the two-level systems are indistinguishable in location. It was found that in both nearest neighbour hopping and hyper-hopping systems, the scaling factor for the charging power as a function of $\kappa$, $N$ and $m$ were consistent. 

It is clear that there is a disparity between the present results for the JCH quantum battery and that what was found for the Dicke model in \cite{Ferraro_2018_PRL}. This motivated us to revisit the Dicke model of a quantum battery.  

\section{Dicke quantum batteries}\label{sec:Dicke_theory}
We begin with the generalised Dicke Hamiltonian \cite{Kirton_2019_AQT},
\begin{align}
H_{Dicke}=& \ \ \omega_ca^{\dagger}a +\omega_a \sum_{n=1}^N \sigma^z_n \nonumber\\
&+\frac{\beta}{\sqrt{N}}\sum_{n=1}^N(a\sigma^+_n+a^{\dagger}\sigma^-_n) \nonumber\\
&+\frac{\beta'}{\sqrt{N}}\sum_{n=1}^N(a\sigma^i_n+a^{\dagger}\sigma^+_n), \label{eq:dicke_ham}
\end{align}
where $\beta$ and $\beta‘$ are the coupling between the photon mode and the atomic excitation degree of freedom for the energy conserving interactions. $\beta’$ is the coupling parameter for the interactions that do not conserve excitation number, namely the $a^\dagger\omega^+_j$ term excites the atom and also produces adds a photon to the system. The Dicke model is a special case of equation \eqref{eq:dicke_ham} where $\beta’=\beta$, while the Tavis-Cummings model is the case where $\beta’=0$. It is worthy to note that there the $\frac{1}{\sqrt{N}}$ terms in \eqref{eq:dicke_ham} are included to ensure that at the thermal limit the energy remains bounded. However, in \cite{Ferraro_2018_PRL} the factor of $\sqrt{N}$ was not included in their Hamiltonian and went on to find that the maximum power scaled according to the relation
\begin{equation}
P_{max}\propto N^{3/2}.\nonumber
\end{equation}
With a power scaling increasing faster than the size of the battery increases, significant interest has been shown in the exciting potential of super charging quantum batteries. The improved charging capabilities has been attributed to entangled states and in the results of the paper we will elucidate the relationship that, the size of the Dicke and JCH quantum batteries in terms of the number of photons and atoms, and the coupling strength, has on the charging power. 

The process of setting up the matrix for the Dicke Hamiltonian is well described in \cite{Ferraro_2018_PRL}, and will be briefly re-iterated here. For a particular state and a given number of two-level systems, $N$, in the cavity there with $q$ of them in the ground state, and $n$ photons. The ground state, where there are $m=n/N=1$ photons per two-level system and $q=N$ of the two-level systems which are in the ground state, can be written 

\begin{equation}
\ket{\psi^N(0)} = \ket{ N,\frac{N}{2},-\frac{N}{2}}.
\end{equation}
Using this, the elements of the matrix for the Dicke Hamiltonian can found from
\begin{align}
&\braket{n',\frac{N}{2},\frac{N}{2}-q' | H^N(t)| n,\frac{N}{2},\frac{N}{2} -q } =\nonumber \\
&\omega_c\big[(n+\frac{N}{2}-q)\delta_{n',n}\delta_{q',q} 
+ \frac{\beta}{\sqrt{N}} \big\{ f^{(1)}_{n,\frac{N}{2},\frac{N}{2}-q} \delta_{n',n+1}\delta_{q',q+1} \nonumber\\
& +f^{(2)}_{n,\frac{N}{2},\frac{N}{2}-q} \delta_{n',n+1}\delta_{q',q-1}
  +f^{(3)}_{n,\frac{N}{2},\frac{N}{2}-q} \delta_{n',n-1}\delta_{q',q+1}\nonumber\\
& +f^{(4)}_{n,\frac{N}{2},\frac{N}{2}-q} \delta_{n',n-1}\delta_{q',q-1} \big\} \big] \label{eq:dicke_mat}
 \end{align}
with the primed perms denoting the final quantities and
\begin{align}
f^{(1)}_{k,j,m} = &\sqrt{ (k+1)[j(j+1)-m(m-1)]},\nonumber \\
f^{(2)}_{k,j,m}=&\sqrt{(k+1)[j(j+1)-m(m+1)]},\nonumber\\
f^{(3)}_{k,j,m}=&\sqrt{k[j(j+1)-m(m-1)]},\nonumber\\
f^{(4)}_{k,j,m}=&\sqrt{k[j(j+1)-m(m+1)]}.\nonumber   \label{eq:dick_f_terms}
\end{align}
From the obtained Hamiltonian the time dependent energy function and the power of the Dicke quantum battery can be determined in the same way that is was in Section \ref{sec:jch_theory}, using equations \eqref{eq:JCH_energy} \& \eqref{eq:pmax}.

\section{Results: Dicke model}\label{sec:dicke_results}
\begin{figure}[h]
	\centering
	\includegraphics[width=0.85\linewidth]{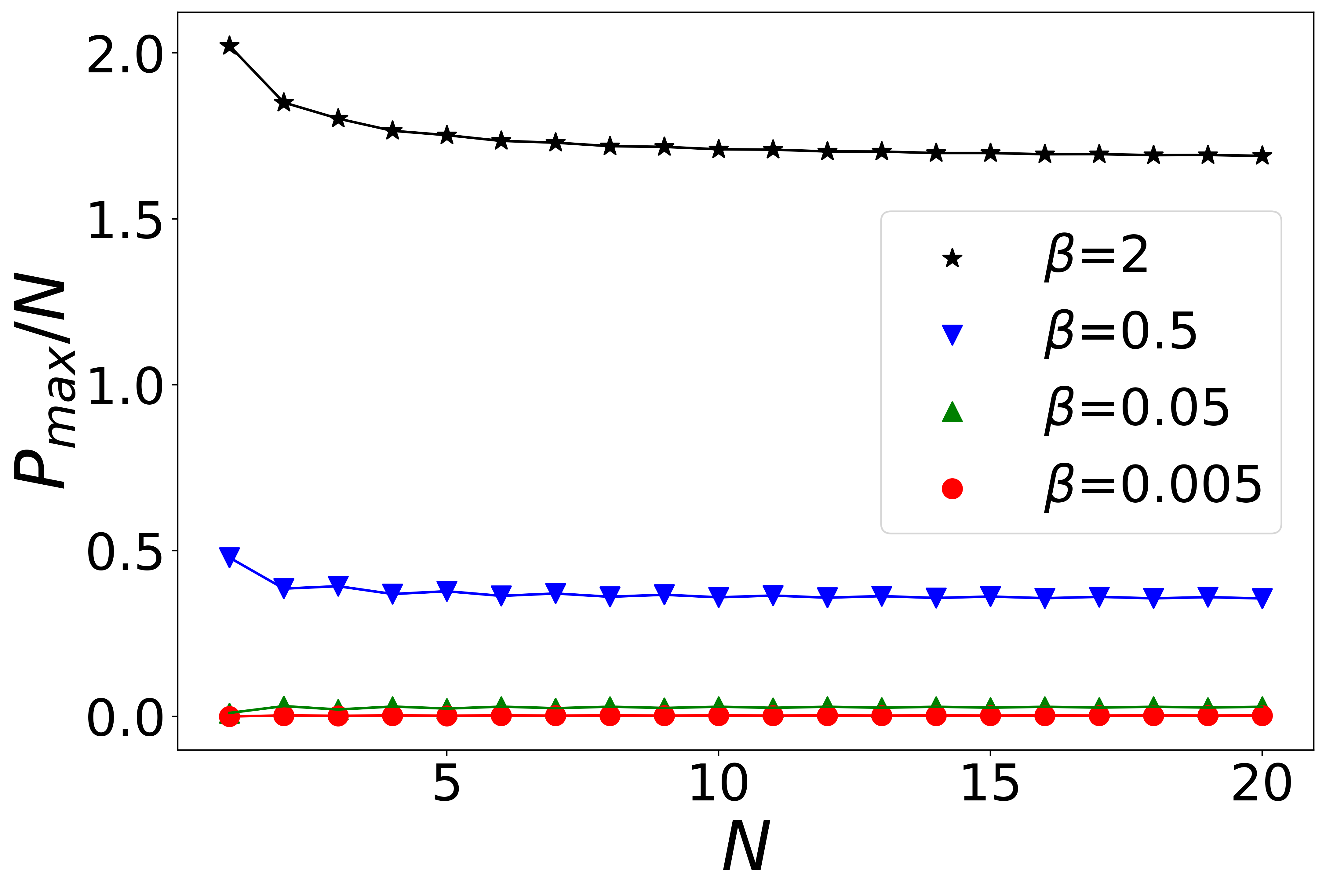}
	\caption{ The power scaling of a Dicke quantum battery for an increasing battery size. The number of two-level systems ranges from 2 to 20, while the photon-atom coupling parameter $\beta$ varies from low coupling to where it dominates the dynamics of the system ($0$ to $2$). Asymptotic behaviour is seen for the scaled power towards a constant value for increasing $N$. } 
	\label{fig:dicke_N_1}
\end{figure}
Initially importance was placed on the model being able to replicate the previous results of \cite{Ferraro_2018_PRL}, which was possible by using the Hamiltonian referenced in their paper, without the factor of $1/\sqrt{N}$. However, when using the  Dicke Hamiltonian, of equation \eqref{eq:dicke_ham}, with the $1/\sqrt{N}$ term in the photon to two-level system coupling terms, the super-charging is not present. This strong agreement between the JCH and the Dicke models, confirms that while the charging power of quantum batteries does increase as the size of the battery increases, it does so by a factor of $N$, i.e. 
\begin{equation}
P_{max} \propto N, \nonumber
\end{equation}
as demonstrated in Fig.\ref{fig:dicke_N_1}. Here, $\beta$ takes on the same values ($\beta = 0,0.05,0.5,2$). While the starting number of photons in the system is taken to be $N$, the Dicke model allows for behaviour that does not conserve the particle number. As a result, to compute the Dicke model limitations on the maximum number of considered photons needs to be placed. In this work we considered systems of a range of photons from $1$ to $5N$ for each data point. $5N$ was taken to be the maximum because good convergence was already found for $4N$. 
\begin{figure}[h]
	\centering
	\includegraphics[width=0.85\linewidth]{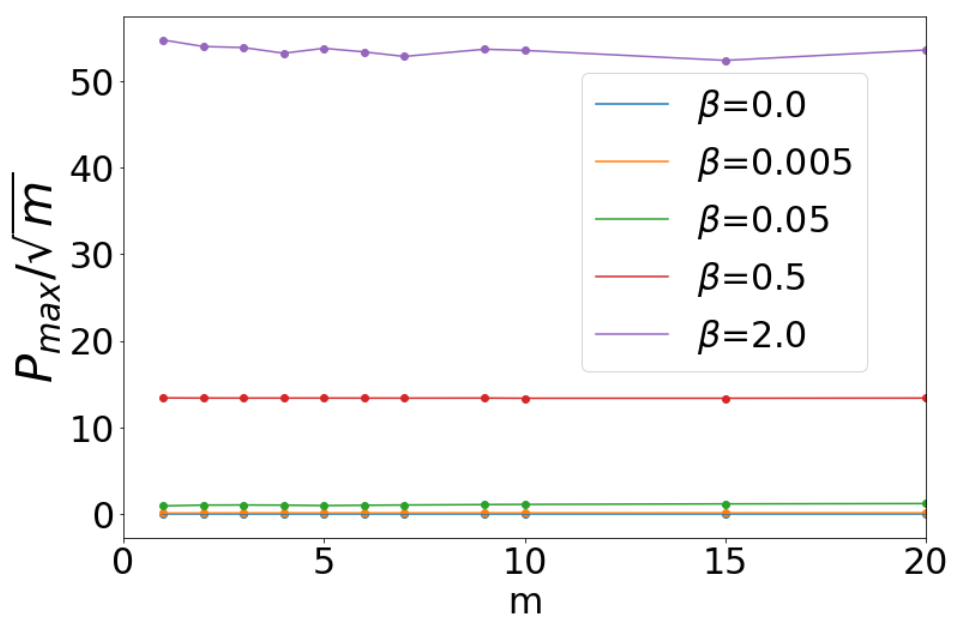}
	\caption{The maximum power as a function of average number of photons $m$, scaled by $\frac{1}{\sqrt{m}}$, for $N=10$ two-level systems.  The legend displays the corresponding values of $\beta$ for each plot, with lines drawn between data points to help visualise the data. } 
	\label{fig:dicke_m}
\end{figure}

Exploring the effect that the average number of photons per two-level system has on $P_{max}$, Fig.\ref{fig:dicke_m} displays that $P_{max}/\sqrt{m}$ converges to a constant value as $m$ increased. This result is the same result which was found in section \ref{sec:JCHresults}, for the JCH battery. The strong agreement between the Dicke and JCH models for a quantum battery is interesting, because of the different ways each model allows the two-level systems to interact with the photon fields. In the Dicke model all of the two-level systems are able to interact with all of the photons at all times, because they exist within the same cavity, while in the JCH model the two-level systems can only interact with the photons in there cavity. This result implies that for large enough $m$ the power returns scale the same way whether you localise the two-level systems or not, without ever considering the strength of the entanglement of states. 

As $m$ gets very large we enter into a regime currently access able by experiments, eg. \cite{Quach_2022_SA}, and when looking at a regime of $m=200$ we found the same scaling relation, implying that the charging power of the Dicke quantum battery is only limited by the number of photons input into the cavity. 
\section{conclusions}
In this paper we have considered the charging quench dynamics of both the JCH and Dicke models. For the JCH model we have found that $P_{max}$ scales linearly with $N$, i.e. there is no quantum advantage in such a system. More generally we also find that as the coupling between the cavities $\kappa$, in the JCH model, is increased $P_{max}$ is reduced. However, there is an increase in  $P_{max}$ when the number of photons, $m$, in each cavity, at $t=0$,  is increased. In this case $P_{max}\propto \sqrt{m}$.

This investigation into the JCH system lead us to revisit the charging quench dynamics of the Dicke model. Starting from a form of the Dicke Hamiltonian which ensures a consistent thermodynamic limit we fiund again that $P_{max}$ scales linearly with the number of the two-level systems in the Dicke cavity. Additionally, we recovered a scaling of $P_{max}\propto \sqrt{m}$, where $m$ is the number of photons per two-level system, at $t=0$, in the Dicke cavity.

\begin{acknowledgments}
Andrew R. Hogan is  supported  by  an  Australian Government Research Training Program Scholarship and by the University of Melbourne.
\end{acknowledgments}

\bibliography{jch_dicke_quantum_batteries_bib.bib}

\begin{thebibliography}{43}%
\makeatletter
\providecommand \@ifxundefined [1]{%
 \@ifx{#1\undefined}
}%
\providecommand \@ifnum [1]{%
 \ifnum #1\expandafter \@firstoftwo
 \else \expandafter \@secondoftwo
 \fi
}%
\providecommand \@ifx [1]{%
 \ifx #1\expandafter \@firstoftwo
 \else \expandafter \@secondoftwo
 \fi
}%
\providecommand \natexlab [1]{#1}%
\providecommand \enquote  [1]{``#1''}%
\providecommand \bibnamefont  [1]{#1}%
\providecommand \bibfnamefont [1]{#1}%
\providecommand \citenamefont [1]{#1}%
\providecommand \href@noop [0]{\@secondoftwo}%
\providecommand \href [0]{\begingroup \@sanitize@url \@href}%
\providecommand \@href[1]{\@@startlink{#1}\@@href}%
\providecommand \@@href[1]{\endgroup#1\@@endlink}%
\providecommand \@sanitize@url [0]{\catcode `\\12\catcode `\$12\catcode
  `\&12\catcode `\#12\catcode `\^12\catcode `\_12\catcode `\%12\relax}%
\providecommand \@@startlink[1]{}%
\providecommand \@@endlink[0]{}%
\providecommand \url  [0]{\begingroup\@sanitize@url \@url }%
\providecommand \@url [1]{\endgroup\@href {#1}{\urlprefix }}%
\providecommand \urlprefix  [0]{URL }%
\providecommand \Eprint [0]{\href }%
\providecommand \doibase [0]{https://doi.org/}%
\providecommand \selectlanguage [0]{\@gobble}%
\providecommand \bibinfo  [0]{\@secondoftwo}%
\providecommand \bibfield  [0]{\@secondoftwo}%
\providecommand \translation [1]{[#1]}%
\providecommand \BibitemOpen [0]{}%
\providecommand \bibitemStop [0]{}%
\providecommand \bibitemNoStop [0]{.\EOS\space}%
\providecommand \EOS [0]{\spacefactor3000\relax}%
\providecommand \BibitemShut  [1]{\csname bibitem#1\endcsname}%
\let\auto@bib@innerbib\@empty
\bibitem [{\citenamefont {Quach}\ \emph {et~al.}(2022)\citenamefont {Quach},
  \citenamefont {McGhee}, \citenamefont {Ganzer}, \citenamefont {Rouse},
  \citenamefont {Lovett}, \citenamefont {Gauger}, \citenamefont {Keeling},
  \citenamefont {Cerullo}, \citenamefont {Lidzey},\ and\ \citenamefont
  {Virgili}}]{Quach_2022_SA}%
  \BibitemOpen
  \bibfield  {author} {\bibinfo {author} {\bibfnamefont {J.~Q.}\ \bibnamefont
  {Quach}}, \bibinfo {author} {\bibfnamefont {K.~E.}\ \bibnamefont {McGhee}},
  \bibinfo {author} {\bibfnamefont {L.}~\bibnamefont {Ganzer}}, \bibinfo
  {author} {\bibfnamefont {D.~M.}\ \bibnamefont {Rouse}}, \bibinfo {author}
  {\bibfnamefont {B.~W.}\ \bibnamefont {Lovett}}, \bibinfo {author}
  {\bibfnamefont {E.~M.}\ \bibnamefont {Gauger}}, \bibinfo {author}
  {\bibfnamefont {J.}~\bibnamefont {Keeling}}, \bibinfo {author} {\bibfnamefont
  {G.}~\bibnamefont {Cerullo}}, \bibinfo {author} {\bibfnamefont {D.~G.}\
  \bibnamefont {Lidzey}},\ and\ \bibinfo {author} {\bibfnamefont
  {T.}~\bibnamefont {Virgili}},\ }\href
  {https://doi.org/10.1126/sciadv.abk3160} {\bibfield  {journal} {\bibinfo
  {journal} {Science Advances}\ }\textbf {\bibinfo {volume} {8}},\ \bibinfo
  {pages} {eabk3160} (\bibinfo {year} {2022})},\ \Eprint
  {https://arxiv.org/abs/https://www.science.org/doi/pdf/10.1126/sciadv.abk3160}
  {https://www.science.org/doi/pdf/10.1126/sciadv.abk3160} \BibitemShut
  {NoStop}%
\bibitem [{\citenamefont {Gemme}\ \emph
  {et~al.}(2022{\natexlab{a}})\citenamefont {Gemme}, \citenamefont {Grossi},
  \citenamefont {Ferraro}, \citenamefont {Vallecorsa},\ and\ \citenamefont
  {Sassetti}}]{Gem_2022_IBM}%
  \BibitemOpen
  \bibfield  {author} {\bibinfo {author} {\bibfnamefont {G.}~\bibnamefont
  {Gemme}}, \bibinfo {author} {\bibfnamefont {M.}~\bibnamefont {Grossi}},
  \bibinfo {author} {\bibfnamefont {D.}~\bibnamefont {Ferraro}}, \bibinfo
  {author} {\bibfnamefont {S.}~\bibnamefont {Vallecorsa}},\ and\ \bibinfo
  {author} {\bibfnamefont {M.}~\bibnamefont {Sassetti}},\ }\href
  {https://doi.org/10.48550/ARXIV.2204.10786} {\bibinfo {title} {Ibm quantum
  platforms: a quantum battery perspective}} (\bibinfo {year}
  {2022}{\natexlab{a}})\BibitemShut {NoStop}%
\bibitem [{\citenamefont {Jaramillo}\ \emph {et~al.}(2016)\citenamefont
  {Jaramillo}, \citenamefont {Beau},\ and\ \citenamefont {del
  Campo}}]{Jar_2016_NJP}%
  \BibitemOpen
  \bibfield  {author} {\bibinfo {author} {\bibfnamefont {J.}~\bibnamefont
  {Jaramillo}}, \bibinfo {author} {\bibfnamefont {M.}~\bibnamefont {Beau}},\
  and\ \bibinfo {author} {\bibfnamefont {A.}~\bibnamefont {del Campo}},\ }\href
  {https://doi.org/10.1088/1367-2630/18/7/075019} {\bibfield  {journal}
  {\bibinfo  {journal} {New Journal of Physics}\ }\textbf {\bibinfo {volume}
  {18}},\ \bibinfo {pages} {075019} (\bibinfo {year} {2016})}\BibitemShut
  {NoStop}%
\bibitem [{\citenamefont {Alicki}\ and\ \citenamefont
  {Fannes}(2013)}]{Ali_2013_PRE}%
  \BibitemOpen
  \bibfield  {author} {\bibinfo {author} {\bibfnamefont {R.}~\bibnamefont
  {Alicki}}\ and\ \bibinfo {author} {\bibfnamefont {M.}~\bibnamefont
  {Fannes}},\ }\href {https://doi.org/10.1103/PhysRevE.87.042123} {\bibfield
  {journal} {\bibinfo  {journal} {Phys. Rev. E}\ }\textbf {\bibinfo {volume}
  {87}},\ \bibinfo {pages} {042123} (\bibinfo {year} {2013})}\BibitemShut
  {NoStop}%
\bibitem [{\citenamefont {Binder}\ \emph {et~al.}(2015)\citenamefont {Binder},
  \citenamefont {Vinjanampathy}, \citenamefont {Modi},\ and\ \citenamefont
  {Goold}}]{Bin_2015_NJP}%
  \BibitemOpen
  \bibfield  {author} {\bibinfo {author} {\bibfnamefont {F.~C.}\ \bibnamefont
  {Binder}}, \bibinfo {author} {\bibfnamefont {S.}~\bibnamefont
  {Vinjanampathy}}, \bibinfo {author} {\bibfnamefont {K.}~\bibnamefont
  {Modi}},\ and\ \bibinfo {author} {\bibfnamefont {J.}~\bibnamefont {Goold}},\
  }\href {https://doi.org/10.1088/1367-2630/17/7/075015} {\bibfield  {journal}
  {\bibinfo  {journal} {New Journal of Physics}\ }\textbf {\bibinfo {volume}
  {17}},\ \bibinfo {pages} {075015} (\bibinfo {year} {2015})}\BibitemShut
  {NoStop}%
\bibitem [{\citenamefont {Hovhannisyan}\ \emph
  {et~al.}(2013{\natexlab{a}})\citenamefont {Hovhannisyan}, \citenamefont
  {Perarnau-Llobet}, \citenamefont {Huber},\ and\ \citenamefont
  {Ac\'{\i}n}}]{Hov_2013_APS}%
  \BibitemOpen
  \bibfield  {author} {\bibinfo {author} {\bibfnamefont {K.~V.}\ \bibnamefont
  {Hovhannisyan}}, \bibinfo {author} {\bibfnamefont {M.}~\bibnamefont
  {Perarnau-Llobet}}, \bibinfo {author} {\bibfnamefont {M.}~\bibnamefont
  {Huber}},\ and\ \bibinfo {author} {\bibfnamefont {A.}~\bibnamefont
  {Ac\'{\i}n}},\ }\href {https://doi.org/10.1103/PhysRevLett.111.240401}
  {\bibfield  {journal} {\bibinfo  {journal} {Phys. Rev. Lett.}\ }\textbf
  {\bibinfo {volume} {111}},\ \bibinfo {pages} {240401} (\bibinfo {year}
  {2013}{\natexlab{a}})}\BibitemShut {NoStop}%
\bibitem [{\citenamefont {Konar}\ \emph {et~al.}(2022)\citenamefont {Konar},
  \citenamefont {Lakkaraju}, \citenamefont {Ghosh},\ and\ \citenamefont
  {Sen(De)}}]{Kon_2022_APS}%
  \BibitemOpen
  \bibfield  {author} {\bibinfo {author} {\bibfnamefont {T.~K.}\ \bibnamefont
  {Konar}}, \bibinfo {author} {\bibfnamefont {L.~G.~C.}\ \bibnamefont
  {Lakkaraju}}, \bibinfo {author} {\bibfnamefont {S.}~\bibnamefont {Ghosh}},\
  and\ \bibinfo {author} {\bibfnamefont {A.}~\bibnamefont {Sen(De)}},\ }\href
  {https://doi.org/10.1103/PhysRevA.106.022618} {\bibfield  {journal} {\bibinfo
   {journal} {Phys. Rev. A}\ }\textbf {\bibinfo {volume} {106}},\ \bibinfo
  {pages} {022618} (\bibinfo {year} {2022})}\BibitemShut {NoStop}%
\bibitem [{\citenamefont {Gemme}\ \emph
  {et~al.}(2022{\natexlab{b}})\citenamefont {Gemme}, \citenamefont {Grossi},
  \citenamefont {Ferraro}, \citenamefont {Vallecorsa},\ and\ \citenamefont
  {Sassetti}}]{Gem_2022_BAT}%
  \BibitemOpen
  \bibfield  {author} {\bibinfo {author} {\bibfnamefont {G.}~\bibnamefont
  {Gemme}}, \bibinfo {author} {\bibfnamefont {M.}~\bibnamefont {Grossi}},
  \bibinfo {author} {\bibfnamefont {D.}~\bibnamefont {Ferraro}}, \bibinfo
  {author} {\bibfnamefont {S.}~\bibnamefont {Vallecorsa}},\ and\ \bibinfo
  {author} {\bibfnamefont {M.}~\bibnamefont {Sassetti}},\ }\href
  {https://www.mdpi.com/2313-0105/8/5/43} {\bibfield  {journal} {\bibinfo
  {journal} {Batteries}\ }\textbf {\bibinfo {volume} {8}} (\bibinfo {year}
  {2022}{\natexlab{b}})}\BibitemShut {NoStop}%
\bibitem [{\citenamefont {Santos}(2021{\natexlab{a}})}]{San_2021_APS}%
  \BibitemOpen
  \bibfield  {author} {\bibinfo {author} {\bibfnamefont {A.~C.}\ \bibnamefont
  {Santos}},\ }\href {https://doi.org/10.1103/PhysRevE.103.042118} {\bibfield
  {journal} {\bibinfo  {journal} {Phys. Rev. E}\ }\textbf {\bibinfo {volume}
  {103}},\ \bibinfo {pages} {042118} (\bibinfo {year}
  {2021}{\natexlab{a}})}\BibitemShut {NoStop}%
\bibitem [{\citenamefont {Le}\ \emph {et~al.}(2018)\citenamefont {Le},
  \citenamefont {Levinsen}, \citenamefont {Modi}, \citenamefont {Parish},\ and\
  \citenamefont {Pollock}}]{Lev_2018_APS}%
  \BibitemOpen
  \bibfield  {author} {\bibinfo {author} {\bibfnamefont {T.~P.}\ \bibnamefont
  {Le}}, \bibinfo {author} {\bibfnamefont {J.}~\bibnamefont {Levinsen}},
  \bibinfo {author} {\bibfnamefont {K.}~\bibnamefont {Modi}}, \bibinfo {author}
  {\bibfnamefont {M.~M.}\ \bibnamefont {Parish}},\ and\ \bibinfo {author}
  {\bibfnamefont {F.~A.}\ \bibnamefont {Pollock}},\ }\href
  {https://doi.org/10.1103/PhysRevA.97.022106} {\bibfield  {journal} {\bibinfo
  {journal} {Phys. Rev. A}\ }\textbf {\bibinfo {volume} {97}},\ \bibinfo
  {pages} {022106} (\bibinfo {year} {2018})}\BibitemShut {NoStop}%
\bibitem [{\citenamefont {Dou}\ \emph {et~al.}(2022)\citenamefont {Dou},
  \citenamefont {Lu}, \citenamefont {Wang},\ and\ \citenamefont
  {Sun}}]{Dou_2022_APS}%
  \BibitemOpen
  \bibfield  {author} {\bibinfo {author} {\bibfnamefont {F.-Q.}\ \bibnamefont
  {Dou}}, \bibinfo {author} {\bibfnamefont {Y.-Q.}\ \bibnamefont {Lu}},
  \bibinfo {author} {\bibfnamefont {Y.-J.}\ \bibnamefont {Wang}},\ and\
  \bibinfo {author} {\bibfnamefont {J.-A.}\ \bibnamefont {Sun}},\ }\href
  {https://doi.org/10.1103/PhysRevB.105.115405} {\bibfield  {journal} {\bibinfo
   {journal} {Phys. Rev. B}\ }\textbf {\bibinfo {volume} {105}},\ \bibinfo
  {pages} {115405} (\bibinfo {year} {2022})}\BibitemShut {NoStop}%
\bibitem [{\citenamefont {Crescente}\ \emph {et~al.}(2020)\citenamefont
  {Crescente}, \citenamefont {Carrega}, \citenamefont {Sassetti},\ and\
  \citenamefont {Ferraro}}]{Crescente_2020}%
  \BibitemOpen
  \bibfield  {author} {\bibinfo {author} {\bibfnamefont {A.}~\bibnamefont
  {Crescente}}, \bibinfo {author} {\bibfnamefont {M.}~\bibnamefont {Carrega}},
  \bibinfo {author} {\bibfnamefont {M.}~\bibnamefont {Sassetti}},\ and\
  \bibinfo {author} {\bibfnamefont {D.}~\bibnamefont {Ferraro}},\ }\href
  {https://doi.org/10.1088/1367-2630/ab91fc} {\bibfield  {journal} {\bibinfo
  {journal} {New Journal of Physics}\ }\textbf {\bibinfo {volume} {22}},\
  \bibinfo {pages} {063057} (\bibinfo {year} {2020})}\BibitemShut {NoStop}%
\bibitem [{\citenamefont {Delmonte}\ \emph {et~al.}(2021)\citenamefont
  {Delmonte}, \citenamefont {Crescente}, \citenamefont {Carrega}, \citenamefont
  {Ferraro},\ and\ \citenamefont {Sassetti}}]{e23050612}%
  \BibitemOpen
  \bibfield  {author} {\bibinfo {author} {\bibfnamefont {A.}~\bibnamefont
  {Delmonte}}, \bibinfo {author} {\bibfnamefont {A.}~\bibnamefont {Crescente}},
  \bibinfo {author} {\bibfnamefont {M.}~\bibnamefont {Carrega}}, \bibinfo
  {author} {\bibfnamefont {D.}~\bibnamefont {Ferraro}},\ and\ \bibinfo {author}
  {\bibfnamefont {M.}~\bibnamefont {Sassetti}},\ }\bibfield  {journal}
  {\bibinfo  {journal} {Entropy}\ }\textbf {\bibinfo {volume} {23}},\ \href
  {https://doi.org/10.3390/e23050612} {10.3390/e23050612} (\bibinfo {year}
  {2021})\BibitemShut {NoStop}%
\bibitem [{\citenamefont {Liu}\ \emph {et~al.}(2019)\citenamefont {Liu},
  \citenamefont {Segal},\ and\ \citenamefont
  {Hanna}}]{doi:10.1021/acs.jpcc.9b06373}%
  \BibitemOpen
  \bibfield  {author} {\bibinfo {author} {\bibfnamefont {J.}~\bibnamefont
  {Liu}}, \bibinfo {author} {\bibfnamefont {D.}~\bibnamefont {Segal}},\ and\
  \bibinfo {author} {\bibfnamefont {G.}~\bibnamefont {Hanna}},\ }\href
  {https://doi.org/10.1021/acs.jpcc.9b06373} {\bibfield  {journal} {\bibinfo
  {journal} {The Journal of Physical Chemistry C}\ }\textbf {\bibinfo {volume}
  {123}},\ \bibinfo {pages} {18303} (\bibinfo {year} {2019})}\BibitemShut
  {NoStop}%
\bibitem [{\citenamefont {Juli\`a-Farr\'e}\ \emph {et~al.}(2020)\citenamefont
  {Juli\`a-Farr\'e}, \citenamefont {Salamon}, \citenamefont {Riera},
  \citenamefont {Bera},\ and\ \citenamefont
  {Lewenstein}}]{PhysRevResearch.2.023113}%
  \BibitemOpen
  \bibfield  {author} {\bibinfo {author} {\bibfnamefont {S.}~\bibnamefont
  {Juli\`a-Farr\'e}}, \bibinfo {author} {\bibfnamefont {T.}~\bibnamefont
  {Salamon}}, \bibinfo {author} {\bibfnamefont {A.}~\bibnamefont {Riera}},
  \bibinfo {author} {\bibfnamefont {M.~N.}\ \bibnamefont {Bera}},\ and\
  \bibinfo {author} {\bibfnamefont {M.}~\bibnamefont {Lewenstein}},\ }\href
  {https://doi.org/10.1103/PhysRevResearch.2.023113} {\bibfield  {journal}
  {\bibinfo  {journal} {Phys. Rev. Research}\ }\textbf {\bibinfo {volume}
  {2}},\ \bibinfo {pages} {023113} (\bibinfo {year} {2020})}\BibitemShut
  {NoStop}%
\bibitem [{\citenamefont {Andolina}\ \emph
  {et~al.}(2019{\natexlab{a}})\citenamefont {Andolina}, \citenamefont {Keck},
  \citenamefont {Mari}, \citenamefont {Giovannetti},\ and\ \citenamefont
  {Polini}}]{PhysRevB.99.205437}%
  \BibitemOpen
  \bibfield  {author} {\bibinfo {author} {\bibfnamefont {G.~M.}\ \bibnamefont
  {Andolina}}, \bibinfo {author} {\bibfnamefont {M.}~\bibnamefont {Keck}},
  \bibinfo {author} {\bibfnamefont {A.}~\bibnamefont {Mari}}, \bibinfo {author}
  {\bibfnamefont {V.}~\bibnamefont {Giovannetti}},\ and\ \bibinfo {author}
  {\bibfnamefont {M.}~\bibnamefont {Polini}},\ }\href
  {https://doi.org/10.1103/PhysRevB.99.205437} {\bibfield  {journal} {\bibinfo
  {journal} {Phys. Rev. B}\ }\textbf {\bibinfo {volume} {99}},\ \bibinfo
  {pages} {205437} (\bibinfo {year} {2019}{\natexlab{a}})}\BibitemShut
  {NoStop}%
\bibitem [{\citenamefont {Rossini}\ \emph {et~al.}(2019)\citenamefont
  {Rossini}, \citenamefont {Andolina},\ and\ \citenamefont
  {Polini}}]{PhysRevB.100.115142}%
  \BibitemOpen
  \bibfield  {author} {\bibinfo {author} {\bibfnamefont {D.}~\bibnamefont
  {Rossini}}, \bibinfo {author} {\bibfnamefont {G.~M.}\ \bibnamefont
  {Andolina}},\ and\ \bibinfo {author} {\bibfnamefont {M.}~\bibnamefont
  {Polini}},\ }\href {https://doi.org/10.1103/PhysRevB.100.115142} {\bibfield
  {journal} {\bibinfo  {journal} {Phys. Rev. B}\ }\textbf {\bibinfo {volume}
  {100}},\ \bibinfo {pages} {115142} (\bibinfo {year} {2019})}\BibitemShut
  {NoStop}%
\bibitem [{\citenamefont {Cruz}\ \emph {et~al.}(2022)\citenamefont {Cruz},
  \citenamefont {Anka}, \citenamefont {Reis}, \citenamefont {Bachelard},\ and\
  \citenamefont {Santos}}]{Cruz_2022}%
  \BibitemOpen
  \bibfield  {author} {\bibinfo {author} {\bibfnamefont {C.}~\bibnamefont
  {Cruz}}, \bibinfo {author} {\bibfnamefont {M.~F.}\ \bibnamefont {Anka}},
  \bibinfo {author} {\bibfnamefont {M.~S.}\ \bibnamefont {Reis}}, \bibinfo
  {author} {\bibfnamefont {R.}~\bibnamefont {Bachelard}},\ and\ \bibinfo
  {author} {\bibfnamefont {A.~C.}\ \bibnamefont {Santos}},\ }\href
  {https://doi.org/10.1088/2058-9565/ac57f3} {\bibfield  {journal} {\bibinfo
  {journal} {Quantum Science and Technology}\ }\textbf {\bibinfo {volume}
  {7}},\ \bibinfo {pages} {025020} (\bibinfo {year} {2022})}\BibitemShut
  {NoStop}%
\bibitem [{\citenamefont {Hovhannisyan}\ \emph
  {et~al.}(2013{\natexlab{b}})\citenamefont {Hovhannisyan}, \citenamefont
  {Perarnau-Llobet}, \citenamefont {Huber},\ and\ \citenamefont
  {Ac\'{\i}n}}]{PhysRevLett.111.240401}%
  \BibitemOpen
  \bibfield  {author} {\bibinfo {author} {\bibfnamefont {K.~V.}\ \bibnamefont
  {Hovhannisyan}}, \bibinfo {author} {\bibfnamefont {M.}~\bibnamefont
  {Perarnau-Llobet}}, \bibinfo {author} {\bibfnamefont {M.}~\bibnamefont
  {Huber}},\ and\ \bibinfo {author} {\bibfnamefont {A.}~\bibnamefont
  {Ac\'{\i}n}},\ }\href {https://doi.org/10.1103/PhysRevLett.111.240401}
  {\bibfield  {journal} {\bibinfo  {journal} {Phys. Rev. Lett.}\ }\textbf
  {\bibinfo {volume} {111}},\ \bibinfo {pages} {240401} (\bibinfo {year}
  {2013}{\natexlab{b}})}\BibitemShut {NoStop}%
\bibitem [{\citenamefont {Andolina}\ \emph
  {et~al.}(2019{\natexlab{b}})\citenamefont {Andolina}, \citenamefont {Keck},
  \citenamefont {Mari}, \citenamefont {Campisi}, \citenamefont {Giovannetti},\
  and\ \citenamefont {Polini}}]{PhysRevLett.122.047702}%
  \BibitemOpen
  \bibfield  {author} {\bibinfo {author} {\bibfnamefont {G.~M.}\ \bibnamefont
  {Andolina}}, \bibinfo {author} {\bibfnamefont {M.}~\bibnamefont {Keck}},
  \bibinfo {author} {\bibfnamefont {A.}~\bibnamefont {Mari}}, \bibinfo {author}
  {\bibfnamefont {M.}~\bibnamefont {Campisi}}, \bibinfo {author} {\bibfnamefont
  {V.}~\bibnamefont {Giovannetti}},\ and\ \bibinfo {author} {\bibfnamefont
  {M.}~\bibnamefont {Polini}},\ }\href
  {https://doi.org/10.1103/PhysRevLett.122.047702} {\bibfield  {journal}
  {\bibinfo  {journal} {Phys. Rev. Lett.}\ }\textbf {\bibinfo {volume} {122}},\
  \bibinfo {pages} {047702} (\bibinfo {year} {2019}{\natexlab{b}})}\BibitemShut
  {NoStop}%
\bibitem [{\citenamefont {Caravelli}\ \emph {et~al.}(2020)\citenamefont
  {Caravelli}, \citenamefont {Coulter-De~Wit}, \citenamefont
  {Garc\'{\i}a-Pintos},\ and\ \citenamefont
  {Hamma}}]{PhysRevResearch.2.023095}%
  \BibitemOpen
  \bibfield  {author} {\bibinfo {author} {\bibfnamefont {F.}~\bibnamefont
  {Caravelli}}, \bibinfo {author} {\bibfnamefont {G.}~\bibnamefont
  {Coulter-De~Wit}}, \bibinfo {author} {\bibfnamefont {L.~P.}\ \bibnamefont
  {Garc\'{\i}a-Pintos}},\ and\ \bibinfo {author} {\bibfnamefont
  {A.}~\bibnamefont {Hamma}},\ }\href
  {https://doi.org/10.1103/PhysRevResearch.2.023095} {\bibfield  {journal}
  {\bibinfo  {journal} {Phys. Rev. Research}\ }\textbf {\bibinfo {volume}
  {2}},\ \bibinfo {pages} {023095} (\bibinfo {year} {2020})}\BibitemShut
  {NoStop}%
\bibitem [{\citenamefont {Zhao}\ \emph {et~al.}(2021)\citenamefont {Zhao},
  \citenamefont {Dou},\ and\ \citenamefont {Zhao}}]{PhysRevA.103.033715}%
  \BibitemOpen
  \bibfield  {author} {\bibinfo {author} {\bibfnamefont {F.}~\bibnamefont
  {Zhao}}, \bibinfo {author} {\bibfnamefont {F.-Q.}\ \bibnamefont {Dou}},\ and\
  \bibinfo {author} {\bibfnamefont {Q.}~\bibnamefont {Zhao}},\ }\href
  {https://doi.org/10.1103/PhysRevA.103.033715} {\bibfield  {journal} {\bibinfo
   {journal} {Phys. Rev. A}\ }\textbf {\bibinfo {volume} {103}},\ \bibinfo
  {pages} {033715} (\bibinfo {year} {2021})}\BibitemShut {NoStop}%
\bibitem [{\citenamefont {Lu}\ \emph {et~al.}(2021)\citenamefont {Lu},
  \citenamefont {Chen}, \citenamefont {Kuang},\ and\ \citenamefont
  {Wang}}]{PhysRevA.104.043706}%
  \BibitemOpen
  \bibfield  {author} {\bibinfo {author} {\bibfnamefont {W.}~\bibnamefont
  {Lu}}, \bibinfo {author} {\bibfnamefont {J.}~\bibnamefont {Chen}}, \bibinfo
  {author} {\bibfnamefont {L.-M.}\ \bibnamefont {Kuang}},\ and\ \bibinfo
  {author} {\bibfnamefont {X.}~\bibnamefont {Wang}},\ }\href
  {https://doi.org/10.1103/PhysRevA.104.043706} {\bibfield  {journal} {\bibinfo
   {journal} {Phys. Rev. A}\ }\textbf {\bibinfo {volume} {104}},\ \bibinfo
  {pages} {043706} (\bibinfo {year} {2021})}\BibitemShut {NoStop}%
\bibitem [{\citenamefont {Sen}\ and\ \citenamefont
  {Sen}(2021)}]{PhysRevA.104.L030402}%
  \BibitemOpen
  \bibfield  {author} {\bibinfo {author} {\bibfnamefont {K.}~\bibnamefont
  {Sen}}\ and\ \bibinfo {author} {\bibfnamefont {U.}~\bibnamefont {Sen}},\
  }\href {https://doi.org/10.1103/PhysRevA.104.L030402} {\bibfield  {journal}
  {\bibinfo  {journal} {Phys. Rev. A}\ }\textbf {\bibinfo {volume} {104}},\
  \bibinfo {pages} {L030402} (\bibinfo {year} {2021})}\BibitemShut {NoStop}%
\bibitem [{\citenamefont {Uzdin}\ \emph {et~al.}(2015)\citenamefont {Uzdin},
  \citenamefont {Levy},\ and\ \citenamefont {Kosloff}}]{Usdin_2015_PRX}%
  \BibitemOpen
  \bibfield  {author} {\bibinfo {author} {\bibfnamefont {R.}~\bibnamefont
  {Uzdin}}, \bibinfo {author} {\bibfnamefont {A.}~\bibnamefont {Levy}},\ and\
  \bibinfo {author} {\bibfnamefont {R.}~\bibnamefont {Kosloff}},\ }\href
  {https://doi.org/10.1103/PhysRevX.5.031044} {\bibfield  {journal} {\bibinfo
  {journal} {Phys. Rev. X}\ }\textbf {\bibinfo {volume} {5}},\ \bibinfo {pages}
  {031044} (\bibinfo {year} {2015})}\BibitemShut {NoStop}%
\bibitem [{\citenamefont {Bhattacharjee}\ and\ \citenamefont
  {Dutta}(2021)}]{Bha_2021_PJB}%
  \BibitemOpen
  \bibfield  {author} {\bibinfo {author} {\bibfnamefont {S.}~\bibnamefont
  {Bhattacharjee}}\ and\ \bibinfo {author} {\bibfnamefont {A.}~\bibnamefont
  {Dutta}},\ }\bibfield  {journal} {\bibinfo  {journal} {Eur. Phys. J. B 94,
  239}\ }\href
  {https://doi.org/https://doi.org/10.1140/epjb/s10051-021-00235-3}
  {https://doi.org/10.1140/epjb/s10051-021-00235-3} (\bibinfo {year}
  {2021})\BibitemShut {NoStop}%
\bibitem [{\citenamefont {Skrzypczyk}\ \emph {et~al.}(2014)\citenamefont
  {Skrzypczyk}, \citenamefont {Short},\ and\ \citenamefont
  {Popescu}}]{Skrzypczyk2014}%
  \BibitemOpen
  \bibfield  {author} {\bibinfo {author} {\bibfnamefont {P.}~\bibnamefont
  {Skrzypczyk}}, \bibinfo {author} {\bibfnamefont {A.~J.}\ \bibnamefont
  {Short}},\ and\ \bibinfo {author} {\bibfnamefont {S.}~\bibnamefont
  {Popescu}},\ }\href {https://doi.org/10.1038/ncomms5185} {\bibfield
  {journal} {\bibinfo  {journal} {Nature Communications}\ }\textbf {\bibinfo
  {volume} {5}},\ \bibinfo {pages} {4185} (\bibinfo {year} {2014})}\BibitemShut
  {NoStop}%
\bibitem [{\citenamefont {Zhang}\ and\ \citenamefont
  {blaauboer}(2018)}]{https://doi.org/10.48550/arxiv.1812.10139}%
  \BibitemOpen
  \bibfield  {author} {\bibinfo {author} {\bibfnamefont {X.}~\bibnamefont
  {Zhang}}\ and\ \bibinfo {author} {\bibfnamefont {M.}~\bibnamefont
  {blaauboer}},\ }\href {https://doi.org/10.48550/ARXIV.1812.10139} {\bibinfo
  {title} {Enhanced energy transfer in a dicke quantum battery}} (\bibinfo
  {year} {2018})\BibitemShut {NoStop}%
\bibitem [{\citenamefont {Ferraro}\ \emph {et~al.}(2018)\citenamefont
  {Ferraro}, \citenamefont {Campisi}, \citenamefont {Andolina}, \citenamefont
  {Pellegrini},\ and\ \citenamefont {Polini}}]{Ferraro_2018_PRL}%
  \BibitemOpen
  \bibfield  {author} {\bibinfo {author} {\bibfnamefont {D.}~\bibnamefont
  {Ferraro}}, \bibinfo {author} {\bibfnamefont {M.}~\bibnamefont {Campisi}},
  \bibinfo {author} {\bibfnamefont {G.~M.}\ \bibnamefont {Andolina}}, \bibinfo
  {author} {\bibfnamefont {V.}~\bibnamefont {Pellegrini}},\ and\ \bibinfo
  {author} {\bibfnamefont {M.}~\bibnamefont {Polini}},\ }\href
  {https://doi.org/10.1103/PhysRevLett.120.117702} {\bibfield  {journal}
  {\bibinfo  {journal} {Phys. Rev. Lett.}\ }\textbf {\bibinfo {volume} {120}},\
  \bibinfo {pages} {117702} (\bibinfo {year} {2018})}\BibitemShut {NoStop}%
\bibitem [{\citenamefont {Hu}\ \emph {et~al.}(2021)\citenamefont {Hu},
  \citenamefont {Qiu}, \citenamefont {Souza}, \citenamefont {Yuan},
  \citenamefont {Zhou}, \citenamefont {Zhang}, \citenamefont {Chu},
  \citenamefont {Pan}, \citenamefont {Hu}, \citenamefont {Li}, \citenamefont
  {Xu}, \citenamefont {Zhong}, \citenamefont {Liu}, \citenamefont {Yan},
  \citenamefont {Tan}, \citenamefont {Bachelard}, \citenamefont {Villas-Boas},
  \citenamefont {Santos},\ and\ \citenamefont {Yu}}]{Hu_2021_arx}%
  \BibitemOpen
  \bibfield  {author} {\bibinfo {author} {\bibfnamefont {C.-K.}\ \bibnamefont
  {Hu}}, \bibinfo {author} {\bibfnamefont {J.}~\bibnamefont {Qiu}}, \bibinfo
  {author} {\bibfnamefont {P.~J.~P.}\ \bibnamefont {Souza}}, \bibinfo {author}
  {\bibfnamefont {J.}~\bibnamefont {Yuan}}, \bibinfo {author} {\bibfnamefont
  {Y.}~\bibnamefont {Zhou}}, \bibinfo {author} {\bibfnamefont {L.}~\bibnamefont
  {Zhang}}, \bibinfo {author} {\bibfnamefont {J.}~\bibnamefont {Chu}}, \bibinfo
  {author} {\bibfnamefont {X.}~\bibnamefont {Pan}}, \bibinfo {author}
  {\bibfnamefont {L.}~\bibnamefont {Hu}}, \bibinfo {author} {\bibfnamefont
  {J.}~\bibnamefont {Li}}, \bibinfo {author} {\bibfnamefont {Y.}~\bibnamefont
  {Xu}}, \bibinfo {author} {\bibfnamefont {Y.}~\bibnamefont {Zhong}}, \bibinfo
  {author} {\bibfnamefont {S.}~\bibnamefont {Liu}}, \bibinfo {author}
  {\bibfnamefont {F.}~\bibnamefont {Yan}}, \bibinfo {author} {\bibfnamefont
  {D.}~\bibnamefont {Tan}}, \bibinfo {author} {\bibfnamefont {R.}~\bibnamefont
  {Bachelard}}, \bibinfo {author} {\bibfnamefont {C.~J.}\ \bibnamefont
  {Villas-Boas}}, \bibinfo {author} {\bibfnamefont {A.~C.}\ \bibnamefont
  {Santos}},\ and\ \bibinfo {author} {\bibfnamefont {D.}~\bibnamefont {Yu}},\
  }\href {https://doi.org/10.48550/ARXIV.2108.04298} {\bibinfo {title} {Optimal
  charging of a superconducting quantum battery}} (\bibinfo {year}
  {2021})\BibitemShut {NoStop}%
\bibitem [{\citenamefont {Hu}\ \emph {et~al.}(2022)\citenamefont {Hu},
  \citenamefont {Qiu}, \citenamefont {Souza}, \citenamefont {Yuan},
  \citenamefont {Zhou}, \citenamefont {Zhang}, \citenamefont {Chu},
  \citenamefont {Pan}, \citenamefont {Hu}, \citenamefont {Li}, \citenamefont
  {Xu}, \citenamefont {Zhong}, \citenamefont {Liu}, \citenamefont {Yan},
  \citenamefont {Tan}, \citenamefont {Bachelard}, \citenamefont {Villas-Boas},
  \citenamefont {Santos},\ and\ \citenamefont {Yu}}]{Hu_2022_IOP}%
  \BibitemOpen
  \bibfield  {author} {\bibinfo {author} {\bibfnamefont {C.-K.}\ \bibnamefont
  {Hu}}, \bibinfo {author} {\bibfnamefont {J.}~\bibnamefont {Qiu}}, \bibinfo
  {author} {\bibfnamefont {P.~J.~P.}\ \bibnamefont {Souza}}, \bibinfo {author}
  {\bibfnamefont {J.}~\bibnamefont {Yuan}}, \bibinfo {author} {\bibfnamefont
  {Y.}~\bibnamefont {Zhou}}, \bibinfo {author} {\bibfnamefont {L.}~\bibnamefont
  {Zhang}}, \bibinfo {author} {\bibfnamefont {J.}~\bibnamefont {Chu}}, \bibinfo
  {author} {\bibfnamefont {X.}~\bibnamefont {Pan}}, \bibinfo {author}
  {\bibfnamefont {L.}~\bibnamefont {Hu}}, \bibinfo {author} {\bibfnamefont
  {J.}~\bibnamefont {Li}}, \bibinfo {author} {\bibfnamefont {Y.}~\bibnamefont
  {Xu}}, \bibinfo {author} {\bibfnamefont {Y.}~\bibnamefont {Zhong}}, \bibinfo
  {author} {\bibfnamefont {S.}~\bibnamefont {Liu}}, \bibinfo {author}
  {\bibfnamefont {F.}~\bibnamefont {Yan}}, \bibinfo {author} {\bibfnamefont
  {D.}~\bibnamefont {Tan}}, \bibinfo {author} {\bibfnamefont {R.}~\bibnamefont
  {Bachelard}}, \bibinfo {author} {\bibfnamefont {C.~J.}\ \bibnamefont
  {Villas-Boas}}, \bibinfo {author} {\bibfnamefont {A.~C.}\ \bibnamefont
  {Santos}},\ and\ \bibinfo {author} {\bibfnamefont {D.}~\bibnamefont {Yu}},\
  }\href {https://doi.org/10.1088/2058-9565/ac8444} {\bibfield  {journal}
  {\bibinfo  {journal} {Quantum Science and Technology}\ }\textbf {\bibinfo
  {volume} {7}},\ \bibinfo {pages} {045018} (\bibinfo {year}
  {2022})}\BibitemShut {NoStop}%
\bibitem [{\citenamefont {Campaioli}\ \emph {et~al.}(2017)\citenamefont
  {Campaioli}, \citenamefont {Pollock}, \citenamefont {Binder}, \citenamefont
  {C\'eleri}, \citenamefont {Goold}, \citenamefont {Vinjanampathy},\ and\
  \citenamefont {Modi}}]{Cam_2017_APS}%
  \BibitemOpen
  \bibfield  {author} {\bibinfo {author} {\bibfnamefont {F.}~\bibnamefont
  {Campaioli}}, \bibinfo {author} {\bibfnamefont {F.~A.}\ \bibnamefont
  {Pollock}}, \bibinfo {author} {\bibfnamefont {F.~C.}\ \bibnamefont {Binder}},
  \bibinfo {author} {\bibfnamefont {L.}~\bibnamefont {C\'eleri}}, \bibinfo
  {author} {\bibfnamefont {J.}~\bibnamefont {Goold}}, \bibinfo {author}
  {\bibfnamefont {S.}~\bibnamefont {Vinjanampathy}},\ and\ \bibinfo {author}
  {\bibfnamefont {K.}~\bibnamefont {Modi}},\ }\href
  {https://doi.org/10.1103/PhysRevLett.118.150601} {\bibfield  {journal}
  {\bibinfo  {journal} {Phys. Rev. Lett.}\ }\textbf {\bibinfo {volume} {118}},\
  \bibinfo {pages} {150601} (\bibinfo {year} {2017})}\BibitemShut {NoStop}%
\bibitem [{\citenamefont {Andolina}\ \emph {et~al.}(2018)\citenamefont
  {Andolina}, \citenamefont {Farina}, \citenamefont {Mari}, \citenamefont
  {Pellegrini}, \citenamefont {Giovannetti},\ and\ \citenamefont
  {Polini}}]{PhysRevB.98.205423}%
  \BibitemOpen
  \bibfield  {author} {\bibinfo {author} {\bibfnamefont {G.~M.}\ \bibnamefont
  {Andolina}}, \bibinfo {author} {\bibfnamefont {D.}~\bibnamefont {Farina}},
  \bibinfo {author} {\bibfnamefont {A.}~\bibnamefont {Mari}}, \bibinfo {author}
  {\bibfnamefont {V.}~\bibnamefont {Pellegrini}}, \bibinfo {author}
  {\bibfnamefont {V.}~\bibnamefont {Giovannetti}},\ and\ \bibinfo {author}
  {\bibfnamefont {M.}~\bibnamefont {Polini}},\ }\href
  {https://doi.org/10.1103/PhysRevB.98.205423} {\bibfield  {journal} {\bibinfo
  {journal} {Phys. Rev. B}\ }\textbf {\bibinfo {volume} {98}},\ \bibinfo
  {pages} {205423} (\bibinfo {year} {2018})}\BibitemShut {NoStop}%
\bibitem [{\citenamefont {Friis}\ and\ \citenamefont
  {Huber}(2018)}]{Friis2018precisionwork}%
  \BibitemOpen
  \bibfield  {author} {\bibinfo {author} {\bibfnamefont {N.}~\bibnamefont
  {Friis}}\ and\ \bibinfo {author} {\bibfnamefont {M.}~\bibnamefont {Huber}},\
  }\href {https://doi.org/10.22331/q-2018-04-23-61} {\bibfield  {journal}
  {\bibinfo  {journal} {{Quantum}}\ }\textbf {\bibinfo {volume} {2}},\ \bibinfo
  {pages} {61} (\bibinfo {year} {2018})}\BibitemShut {NoStop}%
\bibitem [{\citenamefont {Garc\'{\i}a-Pintos}\ \emph
  {et~al.}(2020)\citenamefont {Garc\'{\i}a-Pintos}, \citenamefont {Hamma},\
  and\ \citenamefont {del Campo}}]{PhysRevLett.125.040601}%
  \BibitemOpen
  \bibfield  {author} {\bibinfo {author} {\bibfnamefont {L.~P.}\ \bibnamefont
  {Garc\'{\i}a-Pintos}}, \bibinfo {author} {\bibfnamefont {A.}~\bibnamefont
  {Hamma}},\ and\ \bibinfo {author} {\bibfnamefont {A.}~\bibnamefont {del
  Campo}},\ }\href {https://doi.org/10.1103/PhysRevLett.125.040601} {\bibfield
  {journal} {\bibinfo  {journal} {Phys. Rev. Lett.}\ }\textbf {\bibinfo
  {volume} {125}},\ \bibinfo {pages} {040601} (\bibinfo {year}
  {2020})}\BibitemShut {NoStop}%
\bibitem [{\citenamefont {Rossini}\ \emph {et~al.}(2020)\citenamefont
  {Rossini}, \citenamefont {Andolina}, \citenamefont {Rosa}, \citenamefont
  {Carrega},\ and\ \citenamefont {Polini}}]{PhysRevLett.125.236402}%
  \BibitemOpen
  \bibfield  {author} {\bibinfo {author} {\bibfnamefont {D.}~\bibnamefont
  {Rossini}}, \bibinfo {author} {\bibfnamefont {G.~M.}\ \bibnamefont
  {Andolina}}, \bibinfo {author} {\bibfnamefont {D.}~\bibnamefont {Rosa}},
  \bibinfo {author} {\bibfnamefont {M.}~\bibnamefont {Carrega}},\ and\ \bibinfo
  {author} {\bibfnamefont {M.}~\bibnamefont {Polini}},\ }\href
  {https://doi.org/10.1103/PhysRevLett.125.236402} {\bibfield  {journal}
  {\bibinfo  {journal} {Phys. Rev. Lett.}\ }\textbf {\bibinfo {volume} {125}},\
  \bibinfo {pages} {236402} (\bibinfo {year} {2020})}\BibitemShut {NoStop}%
\bibitem [{\citenamefont {Zhao}\ \emph {et~al.}(2022)\citenamefont {Zhao},
  \citenamefont {Dou},\ and\ \citenamefont {Zhao}}]{PhysRevResearch.4.013172}%
  \BibitemOpen
  \bibfield  {author} {\bibinfo {author} {\bibfnamefont {F.}~\bibnamefont
  {Zhao}}, \bibinfo {author} {\bibfnamefont {F.-Q.}\ \bibnamefont {Dou}},\ and\
  \bibinfo {author} {\bibfnamefont {Q.}~\bibnamefont {Zhao}},\ }\href
  {https://doi.org/10.1103/PhysRevResearch.4.013172} {\bibfield  {journal}
  {\bibinfo  {journal} {Phys. Rev. Research}\ }\textbf {\bibinfo {volume}
  {4}},\ \bibinfo {pages} {013172} (\bibinfo {year} {2022})}\BibitemShut
  {NoStop}%
\bibitem [{\citenamefont {Zhang}\ \emph {et~al.}(2019)\citenamefont {Zhang},
  \citenamefont {Yang}, \citenamefont {Fu},\ and\ \citenamefont
  {Wang}}]{PhysRevE.99.052106}%
  \BibitemOpen
  \bibfield  {author} {\bibinfo {author} {\bibfnamefont {Y.-Y.}\ \bibnamefont
  {Zhang}}, \bibinfo {author} {\bibfnamefont {T.-R.}\ \bibnamefont {Yang}},
  \bibinfo {author} {\bibfnamefont {L.}~\bibnamefont {Fu}},\ and\ \bibinfo
  {author} {\bibfnamefont {X.}~\bibnamefont {Wang}},\ }\href
  {https://doi.org/10.1103/PhysRevE.99.052106} {\bibfield  {journal} {\bibinfo
  {journal} {Phys. Rev. E}\ }\textbf {\bibinfo {volume} {99}},\ \bibinfo
  {pages} {052106} (\bibinfo {year} {2019})}\BibitemShut {NoStop}%
\bibitem [{\citenamefont {Chang}\ \emph {et~al.}(2021)\citenamefont {Chang},
  \citenamefont {Yang}, \citenamefont {Dong}, \citenamefont {Fu}, \citenamefont
  {Wang},\ and\ \citenamefont {Zhang}}]{Chang_2021}%
  \BibitemOpen
  \bibfield  {author} {\bibinfo {author} {\bibfnamefont {W.}~\bibnamefont
  {Chang}}, \bibinfo {author} {\bibfnamefont {T.-R.}\ \bibnamefont {Yang}},
  \bibinfo {author} {\bibfnamefont {H.}~\bibnamefont {Dong}}, \bibinfo {author}
  {\bibfnamefont {L.}~\bibnamefont {Fu}}, \bibinfo {author} {\bibfnamefont
  {X.}~\bibnamefont {Wang}},\ and\ \bibinfo {author} {\bibfnamefont {Y.-Y.}\
  \bibnamefont {Zhang}},\ }\href {https://doi.org/10.1088/1367-2630/ac2a5b}
  {\bibfield  {journal} {\bibinfo  {journal} {New Journal of Physics}\ }\textbf
  {\bibinfo {volume} {23}},\ \bibinfo {pages} {103026} (\bibinfo {year}
  {2021})}\BibitemShut {NoStop}%
\bibitem [{\citenamefont {Santos}(2021{\natexlab{b}})}]{PhysRevE.103.042118}%
  \BibitemOpen
  \bibfield  {author} {\bibinfo {author} {\bibfnamefont {A.~C.}\ \bibnamefont
  {Santos}},\ }\href {https://doi.org/10.1103/PhysRevE.103.042118} {\bibfield
  {journal} {\bibinfo  {journal} {Phys. Rev. E}\ }\textbf {\bibinfo {volume}
  {103}},\ \bibinfo {pages} {042118} (\bibinfo {year}
  {2021}{\natexlab{b}})}\BibitemShut {NoStop}%
\bibitem [{\citenamefont {Dicke}(1954)}]{Dicke_1954_APS}%
  \BibitemOpen
  \bibfield  {author} {\bibinfo {author} {\bibfnamefont {R.~H.}\ \bibnamefont
  {Dicke}},\ }\href {https://doi.org/10.1103/PhysRev.93.99} {\bibfield
  {journal} {\bibinfo  {journal} {Phys. Rev.}\ }\textbf {\bibinfo {volume}
  {93}},\ \bibinfo {pages} {99} (\bibinfo {year} {1954})}\BibitemShut {NoStop}%
\bibitem [{\citenamefont {Jaynes}\ and\ \citenamefont
  {Cummings}(1963)}]{Jaynes_1963}%
  \BibitemOpen
  \bibfield  {author} {\bibinfo {author} {\bibfnamefont {E.}~\bibnamefont
  {Jaynes}}\ and\ \bibinfo {author} {\bibfnamefont {F.}~\bibnamefont
  {Cummings}},\ }\href {https://doi.org/10.1109/PROC.1963.1664} {\bibfield
  {journal} {\bibinfo  {journal} {Proceedings of the IEEE}\ }\textbf {\bibinfo
  {volume} {51}},\ \bibinfo {pages} {89} (\bibinfo {year} {1963})}\BibitemShut
  {NoStop}%
\bibitem [{\citenamefont {Kirton}\ \emph {et~al.}(2019)\citenamefont {Kirton},
  \citenamefont {Roses}, \citenamefont {Keeling},\ and\ \citenamefont
  {Dalla~Torre}}]{Kirton_2019_AQT}%
  \BibitemOpen
  \bibfield  {author} {\bibinfo {author} {\bibfnamefont {P.}~\bibnamefont
  {Kirton}}, \bibinfo {author} {\bibfnamefont {M.~M.}\ \bibnamefont {Roses}},
  \bibinfo {author} {\bibfnamefont {J.}~\bibnamefont {Keeling}},\ and\ \bibinfo
  {author} {\bibfnamefont {E.~G.}\ \bibnamefont {Dalla~Torre}},\ }\href
  {https://doi.org/https://doi.org/10.1002/qute.201800043} {\bibfield
  {journal} {\bibinfo  {journal} {Advanced Quantum Technologies}\ }\textbf
  {\bibinfo {volume} {2}},\ \bibinfo {pages} {1800043} (\bibinfo {year}
  {2019})},\ \Eprint
  {https://arxiv.org/abs/https://onlinelibrary.wiley.com/doi/pdf/10.1002/qute.201800043}
  {https://onlinelibrary.wiley.com/doi/pdf/10.1002/qute.201800043} \BibitemShut
  {NoStop}%
\end{thebibliography}%

\end{document}